\documentstyle[prb,aps,eqsecnum,twocolumn]{revtex}

\newcommand{\de}{^{o}}

\newcommand{\beq}{\begin{equation}}
\newcommand{\eeq}{\end{equation}}
\newcommand{\bea}{\begin{eqnarray}}
\newcommand{\eea}{\end{eqnarray}}

\newcommand{\bs}{{\bf{S}}}

\newcommand{\bst}{{\bf{L}}}

\newcommand{\etal}{{\em et al.}}

\newcommand{\ns}{N_s}

\def\jour#1#2#3#4{{#1} {\bf #2}, #3 (#4).}
\def\tit#1#2#3#4#5{{#1} {\bf #2}, #3 (#4).}
\def\epl{Europhys. Lett.}
\def\prl{Phys. Rev. Lett.}

\def\prb{Phys. Rev. B}

\def\jpc{J. Phys. C}
\def\jap{J. Appl. Phys.}
\def\zpb{Z. Phys. B}
\def\ssc{Solid State Comm.}
\def\zetp{Zh. Eksp. Teor. Fiz.}
\def\jetp{Sov. Phys. JETP}

\begin{document}
\draft

\twocolumn[\hsize\textwidth\columnwidth\hsize\csname @twocolumnfalse\endcsname

\title{Low-temperature properties of classical, geometrically
frustrated antiferromagnets}
\author{R. Moessner and J. T. Chalker}
\address{Theoretical Physics, Oxford University,
1 Keble Road, Oxford OX1 3NP, UK}
\maketitle


\begin{abstract}
We study the ground-state and low-energy properties of classical
vector spin models with nearest-neighbour antiferromagnetic
interactions on a class of geometrically frustrated lattices
which includes the kagome and pyrochlore lattices.  We explore the
behaviour of these magnets that results from their large ground-state
degeneracies, emphasising universal features and
systematic differences between individual models. We investigate
the circumstances under which thermal
fluctuations select a particular subset of the ground states,
and find that this happens only for the models with the
smallest ground-state degeneracies. For the pyrochlore magnets, we
give an explicit construction of all ground states, and show that
they
are not separated by internal energy barriers.
We study the precessional spin dynamics of the Heisenberg
pyrochlore antiferromagnet.
There is no freezing transition or selection of
preferred states. Instead, the
relaxation time at low temperature, $T$,
is of order $\hbar/k_BT$.
We argue that this behaviour can also be expected in
some
other systems, including
the Heisenberg model for the compound $SrCr_{8}Ga_{4}O_{19}$.
\end{abstract}


]

\input{psfig}

\section{Introduction}
\begin{figure}
\vspace{-1.7cm}
\centerline{\psfig{figure=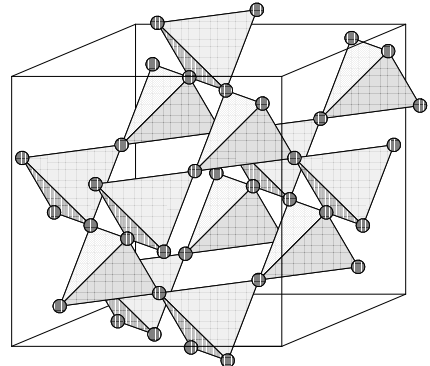,width=7cm}}
\vspace{-4cm}
\caption{The pyrochlore lattice}
\label{fig:pyrochlore}
\end{figure}

Experimental and theoretical studies in recent years have found that
geometrically frustrated antiferromagnets display properties quite
unlike those of other magnetic systems.\cite{reviews} These materials
have magnetic ions located on lattices of site-sharing frustrated
units - usually triangles or tetrahedra.  One of the best-studied
systems in this class is the layered compound $SrCr_{8}Ga_{4}O_{19}$
($SCGO$).\cite{obradors,SCGO,broholm3,SCGOprb,martinez,broholm1,broholm2,schifferscgo,aeppli}
Attention has focussed on the fact that the majority of its magnetic
$Cr^{3+}$-ions reside on the sites kagome lattices, although the full
structure is more complex.  Following the interest in kagome magnets
generated by studies of $SCGO$, a great deal of attention has been
devoted to the oxide and fluoride pyrochlore magnets, in which the
magnetic ions form a lattice of corner-sharing tetrahedra as depicted
in Fig.~\ref{fig:pyrochlore}.  Neutron
scattering\cite{broholm3,martinez,broholm1,broholm2,schifferscgo,aeppli,kurtz,bevaart,gaurei,greedan,harrisneutron1,harrisneutron2,martinmc}
and muon spin relaxation\cite{uemura,dunsiger,muon} experiments on
$SCGO$\ and the pyrochlores have detected only short-range magnetic
correlations and a slowing-down of fluctuations at low
temperatures.\cite{reviews} More generally, it is apparently a
characteristic property of geometrically frustrated magnets that they
do not order at the temperature expected from the magnitude of the
Curie-Weiss constant, $|\Theta_{CW}|$.  Instead they remain in the
paramagnetic phase to a much lower temperature with - typically - spin
freezing at $T_F \ll |\Theta_{CW}|$.\cite{raju,reigre,gingrascrit}

A detailed understanding of the origin of such generic
features has been slow to emerge. Moreover, there has been little work 
to explain systematic differences between individual examples of 
these
magnetic systems. For instance, whereas in the
Heisenberg kagome antiferromagnet thermal fluctuations 
give rise to entropic ground-state
selection,\cite{chalkerkagome,ritchey} known as order by
disorder,\cite{villainobdo,shenderquantum,henleydisorder,henley,long}
this phenomenon appears to be absent for 
some related systems.\cite{rute,husimi}
The reason for the difference is unclear,
as are the general conditions 
under which such selection should be expected for 
geometrically frustrated magnets.
In this context, it is interesting to ask whether $SCGO$
inherits its properties from 
those of the kagome Heisenberg antiferromagnet,
or whether its behaviour is closer to that of
the pyrochlore antiferromagnet, since
an alternative and more complete description of the structure is to
regard a layer of $SCGO$ as a slab
cut from the pyrochlore lattice, consisting of three consecutive
[111] lattice planes.

A further area for investigation, in
addition to the statistical mechanics of geometrically 
frustrated antiferromagnets,
is their low-temperature
dynamics, which has so far
received only limited attention.\cite{reimersmc,keren} 
Dynamical correlations are likely to be profoundly influenced 
by the large ground state degeneracy
of these systems, and constitute one
of their most interesting aspects.

In an attempt to extend understanding of these problems, we have
studied the low-temperature properties of the classical Heisenberg
model with nearest-neighbour interactions on a class of geometrically
frustrated lattices. This description neglects various additional
features -- such as
anisotropy,\cite{fef3,reimersanis,chandra,harrisspinice,relief}
disorder,\cite{origami} dipolar\cite{husedip} or further-neighbour
interactions,\cite{reimers,harrisab} and quantum
effects\cite{ritchey,harrisquantum,zeng,sachdev,chubukov2,lacroix,leche,cana}
-- which can play an important
role in real materials, particularly near and below $T_F$. However, it
may provide a good treatment for the temperature window $T<T_F\ll
|\Theta_{CW}|$, and its simplicity should make it well-suited for
capturing the generic features of these systems,
as well as providing a basis for future investigations incorporating 
additional interactions or quantum fluctuations.

Our work, parts of which have been described in
Refs.~\onlinecite{pyrosh} and \onlinecite{thesis}, concentrates on the
pyrochlore antiferromagnet, but we address several questions in a more
general context.  We start by analysing the origin and extent of the
ground-state degeneracy of geometrically frustrated magnets
(Sect.~\ref{sect:models}). We discuss the nature of the ground-state
manifold of pyrochlore antiferromagnets with $n$-component spins
(Sect.~\ref{chapter:gs}).  We give an explicit construction of all
ground states of these magnets, and show that they are not separated
by energy barriers. Although typical ground states are disordered, we
show that certain correlations remain, which give rise to distinctive
features in magnetic neutron scattering.  We study, both analytically
(Sect.~\ref{sect:singletet}) and numerically
(Sect.~\ref{sect:gsseln}), the existence of order by disorder for a
general class of geometrically frustrated antiferromagnets and find
that it occurs only for magnets with small ground-state
degeneracies. In particular, it is absent from the Heisenberg
pyrochlore magnet, which therefore has neither internal energy nor
large free energy barriers separating different ground states. Because
of this, the system is not trapped near a particular state at low
temperatures.  Our study of the precessional dynamics at low
temperatures and for long times (Sect.~\ref{chapter:dy}) reveals that
the decay of the autocorrelation function is exponential in time, $t$,
with a timescale inversely proportional to the temperature and {\em
independent} of the exchange energy: $\langle {\bf S}_i(0).{\bf
S}_i(t) \rangle = \exp(-ck_BTt/\hbar)$, where $c$ is ${\cal O}(1)$.
In agreement with Reimers' earlier Monte Carlo simulations,
\cite{reimersmc} we find that the spin-freezing transition observed
experimentally does not happen in the simple Heisenberg model we
consider.  We discuss recent experiments on pyrochlore
magnets\cite{harrisneutron2} and $SCGO$\cite{broholm1,aeppli} in the
light of these results.

Since spin correlations are short-ranged in both space and time, the
Heisenberg pyrochlore antiferromagnet can be labelled a classical spin
liquid or, following Villain,\cite{villain} a cooperative paramagnet.


\section{The Heisenberg spin Hamiltonian on geometrically frustrated
lattices} 
\label{sect:models}

Consider
$n$-component classical spins, $\bs_i$, with $|\bs_i|=1$,
arranged in corner-sharing units of $q$\ sites.
Each spin is coupled antiferromagnetically with its $q-1$\ neighbours
in each unit,
so that the Hamiltonian is
\bea
H & = &   
J \sum_{<i,j>} \bs_i \cdot \bs_j 
\equiv 
\frac{J}{2}\sum_{\alpha}|{\bf L}_{\alpha}|^2-~\frac{J}{2}Nq.
\label{eq:spinhamil2}
\eea 
Here, $J$\ is the exchange constant and ${\bf L}_{\alpha}$\ is
total spin in unit $\alpha$. The sum on $\left<i,j \right>$\ runs over
all neighbouring pairs and the sum on $\alpha$ runs over the $N$ units
making up the system. 

Note that our motivation for considering $n$-component spins is to
shed light on the systematics of geometrically frustrated
antiferromagnets.  Because of this, we take the $n$-component
spin-space to be the same at each site. Of course, the case $n=2$ can
also arise physically in a Heisenberg system with easy-plane
anisotropy: in this event, which has been studied in
Ref.~\onlinecite{obdogingras}, the easy planes are orientated
differently at different sites, in accordance with the local symmetry
axes.

An instructive way of thinking about the strength of the geometric
frustration is to consider the extra ground-state degeneracy 
which it gives rise to, in addition to that
stemming from 
the symmetry of
the Hamiltonian. It is this extra degeneracy which lies behind 
many of the
physical properties peculiar to geometrically frustrated systems. 
To determine the number, $D$, of degrees of freedom in the ground state,
we use a Maxwellian counting argument,
\cite{reimers,philmag} and evaluate $D_M \equiv F-K$,
the difference between
the total number, $F$, of degrees of freedom in the system, and the
number, $K$, of constraints that must be imposed
to restrict the system to its ground states.
In general, as discussed below, $D_M \not= D$,
but for pyrochlore 
antiferromagnets
we argue in Sect.~\ref{subsect:heisenberggs}
that $D_M/D \to 1$ as $N \to \infty$.

To evaluate $K$, note that, from Eq.~\ref{eq:spinhamil2}, a 
configuration is a ground state
provided ${\bf L}_{\alpha}=0$ for each unit separately. This imposes
$Nn$\ constraints. To find $F$, we start from the fact that
the number of degrees of freedom  is simply $n-1$ per spin. 
Expressed in terms of the number, $N$, of units, $F$ depends 
on their geometric arrangement.
For corner-sharing units of $q$ spins, $F=Nq(n-1)/2$.
Alternative arrangements generally result in smaller values of $F/N$
and in ground states that are not extensively degenerate.
For example, 
if bonds are shared between units
-- as in the triangular and face-centred cubic lattices for
$q=3$\ and $q=4$\ respectively -- $F$ is lower than if only sites 
are shared -- as in the kagome and the
pyrochlore lattices -- since each spin belongs only to
$b=2$\ units in the latter case but to more ($b=6$\ and $8$,
respectively) in the former. In the general case, we obtain $F=N
q(n-1)/b$. Hence, $D_M/N=\left[ q(n-1)/b-n \right] $. $D_M$\ grows
with $q$ and, for $q>b$, with $n$. In order to obtain $D_M>0$, 
we require $q>b$,
which is the case only for corner-sharing arrangements. The physically
realisable example for which $D_M$\ is maximal is that for which $q$ and
$n$ are both maximal: Heisenberg spins ($n=3$) on the pyrochlore
lattice ($q=4$) represent the only simple system for which $D_M$ is 
positive
and extensive. It is partly for this reason that the pyrochlore
Heisenberg antiferromagnet is particularly interesting.

This counting argument can go wrong in two ways. Firstly, the $K$\
constraints may not be independent, as happens for
Heisenberg spins on the kagome lattice, where
$D_M=0$\ but an extensive ground-state degeneracy nonetheless arises.
Secondly, 
for some lattices there may be no spin configurations that satisfy the conditions ${\bf L}_{\alpha} = 0$ for all $\alpha$.

\begin{figure}
\centerline{\psfig{figure=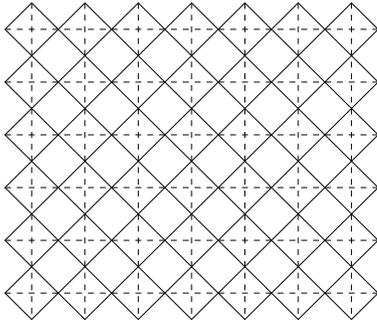,width=5cm}}
\caption{The square lattice with crossings. Both solid and broken
lines denote exchange interactions. Spins reside on the intersections 
of the solid
lines. 
} 
\label{fig:squarecrossing}
\end{figure}

Many of the results presented in this paper do not depend on the
details of the lattice under consideration but rather on the size $q$\
of the corner-sharing units. We find it useful to consider, in
addition to the pyrochlore lattice, the two-dimensional square lattice
with crossings\cite{liebmann} (Fig.~\ref{fig:squarecrossing}), which
is not known to occur in nature but is easy to visualise. Like the
pyrochlore lattice, from which it can be obtained by a projection in a
$\left<001\right>$\ direction, it has $q=4$\, and, 
with Heisenberg spins, $D_M=N$.

\begin{figure}
\centerline{\psfig{figure=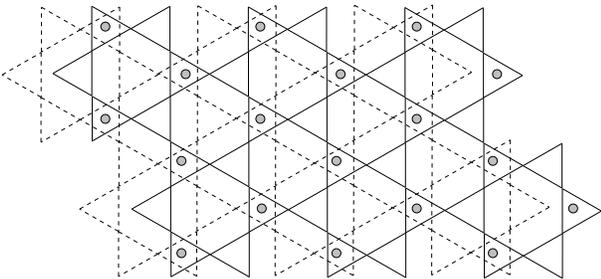,width=8cm}}
\caption{Projection of the sites of the magnetic $Cr^{3+}$~ions
in one layer of
SCGO. The ions occupy sites of the kagome lattices in the
top (solid lines) and bottom (dashed lines) planes. In the middle
plane, the ions are located on a triangular lattice (indicated by
circles). 
All lines denote exchange interactions; 
there are also interactions between a spin in the middle plane
and those in the two triangles which enclose it.
}
\label{fig:scgo}
\end{figure}

Also, more complicated corner-sharing arrangements of frustrated units
are possible. Of particular experimental importance, as mentioned
above, is the combination of triangles and tetrahedra found in $SCGO$,
which is depicted in Fig.~\ref{fig:scgo}.

\section{The ground states of classical antiferromagnets on the
pyrochlore lattice}
\label{chapter:gs}
\label{sect:gscons}

It has been realised for a long time that antiferromagnets on the
pyrochlore lattice have a vast ground-state
degeneracy,\cite{andersonpyro,villain} but no explicit
construction of the ground states has as yet been available. 
The nature of various submanifolds of the ground-state manifold is
however known. These submanifolds are defined by imposing extra
constraints on the spin arrangement, in addition to the requirement
that it be a ground sate.
A simple example is the set of
four-sublattice states, in which the four spins of each unit cell are
arranged to be oriented the same way everywhere. Any four-spin
arrangement that is a ground state for the single tetrahedron (see
section \ref{subsubsect:singletetheis}) yields a ground state for the
entire system by periodic repetition. Of these states, those with 
two spins 
parallel and two antiparallel to a given axis
are the simplest conceivable ones.  Villain\cite{villain} has
described a larger ground-state submanifold for the
Heisenberg model, in which the spins of each tetrahedron form
two antiparallel pairs.
It turns out that for $XY$\ model all ground states are of this kind,
as
described in Sect.~\ref{sect:loops}.

In the following, we present complete constructions of the ground 
states for classical antiferromagnets with $n$-component spins on 
the pyrochlore lattice.  We also show that the
ground-state manifold is connected. We then
examine the conequences of spin correlations in typical ground states for
elastic
neutron scattering. We conclude this section with a
discussion of the nature of the ground-state degrees of freedom
in such magnets.

\subsection{The single tetrahedron}
\label{subsubsect:singletetheis}

\begin{figure}
\centerline{\psfig{figure=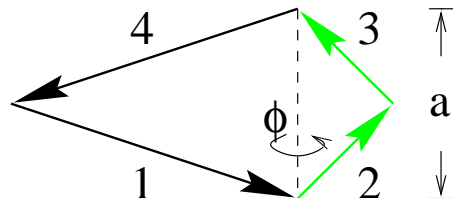,width=6cm}}
\caption{Four spins of equal length with vector sum ${\bf L}=0$.~Spins 1
and 4 lie in the plane of the paper, spins 2 and 3 need not.}
\label{fig:singlegs}
\end{figure}

The ground states of a single tetrahedron
are those states in which the sum, ${\bf L}$, of
the four spin vectors has the value $\bst=0$. In such a configuration,
any two spins enclose the same angle as the other two. 
For Heisenberg spins, these configurations can be parametrised by two
coordinates (e.g. $a$~and $\phi$~in Fig.~\ref{fig:singlegs}). The
crucial feature is that, for any fixed $a$ except $a=0$,
one can choose
$\phi$~independently. In the special case, $a=0$, if spins 1 and 4 are
antiparallel, there are two degrees of freedom associated with the
remaining two spins, while if spins 1 and 4 are
parallel, there is no remaining freedom.
These exceptional states, in which all spins of a tetrahedron  
are collinear,
can play a central role in
determining the thermodynamics of the system
because they are favoured by thermal fluctuations, as discussed
in section \ref{subs:singletet}.

For the $XY$ antiferromagnet, there is only one continuous degree 
of freedom, $a$,
since if spins are coplanar,
$\phi = 0,\pi$.
Ground states are therefore the configurations
with two pairs of antiparallel spins.

\subsection{The Heisenberg antiferromagnet}
\label{subsect:heisenberggs}

\subsubsection{The construction of the ground states}
\label{subsubsect:heisenbergconstruction}

We give in this subsection a stepwise procedure for constructing any
ground state of the Heisenberg pyrochlore antiferromagnet, from which
the number of ground-state degrees of freedom, $D$, can be determined
directly. We also consider a similar proceedure for the square lattice
with crossings since it is essentially the same but easier to explain
and visualise.  The idea in both cases is that the ground state can be
built up by choosing the orientations of spins on successive layers
(planes or lines) of the lattice, in a way that requires no
adjustments of spins in planes or lines already visited.  We consider
systems with open boundary conditions: in the context of this section,
periodic boundary conditions appear to introduce significant
additional mathematical difficulties.

We define a layer, for the square lattice with crossings, to be a [10]
plane (Fig.~\ref{fig:cone}), and for the pyrochlore lattice to be a
[100] plane.  In both cases, a layer contains the spins lying on
equivalent edges of squares or tetrahedra -- referred to as units from
hereon -- which are next-nearest, but not nearest, neighbours, to
other units with spins in the layer. The spins of each unit are shared
between two adjacent layers. Conversely, each spin belongs to a unit
extending above and one extending below the layer.

First, choose the orientations of the spins on the lowest layer
of the lattice. This amounts to choosing a value for
$a$\ in each unit with spins on the bottom layer. 
There are no restrictions on how to
do this. Next, consider the adjacent layer: when choosing the
orientation of spins on
that layer, one has to satisfy the ground-state
condition. For each unit, this leaves
one degree of freedom, $\phi$, except in the special case, $a=0$. 
For this special case, one has to distinguish two situations. If the
bottom spins of a unit are antiparallel, there are two degrees
of freedom when choosing the ground state orientation
of the upper pair of spins: when counting ground-state coordinates,
the loss of a coordinate that follows from
the additional
constraint, $a=0$, is exactly balanced by the gain of an
additional degree of freedom for the upper pair of spins.
In the alternative situation, in which the first pair of spins is parallel,
no freedom remains, and we
obtain a lower-dimensional submanifold of the ground state.

At this stage, we have fixed the orientation of spins in the lowest 
two layers of the lattice.
Repeating the procedure in the subsequent layers,
the value for $a$ in each unit is
determined by earlier choices,
while one degree of freedom, $\phi$, remains
for each unit of the system.

Since any given ground state can be built up (or copied) layer by
layer in this way, the construction can be used to generate all possible
ground states. By this construction we have demonstrated that 
the extensive part of
the dimension of the ground-state manifold is equal to the
number of units in the system.

\subsubsection{The connectedness of the ground-state manifold}
\label{subsubsect:heisenbergconnected}

We show that the ground-state manifold is connected by demonstrating
that any ground state can be continuously deformed into any other
ground state without cost in energy.  To do so, we choose a reference
ground state and give an explicit construction by which the reference
state can be reached from any ground state, without leaving the ground
state manifold. This is done by considering successive layers of the
system, and bringing the spins in each layer into the orientation that
they have in the reference state, using moves within the ground-state
manifold which leave spins unchanged in the layers already visited.
The reference state is chosen to be one in which all spins on lattice
sites equivalent under translation have the same orientation, and in
which each spin is antiparallel to the other spin belonging to the
same layer and unit.

\begin{figure}
\centerline{\psfig{figure=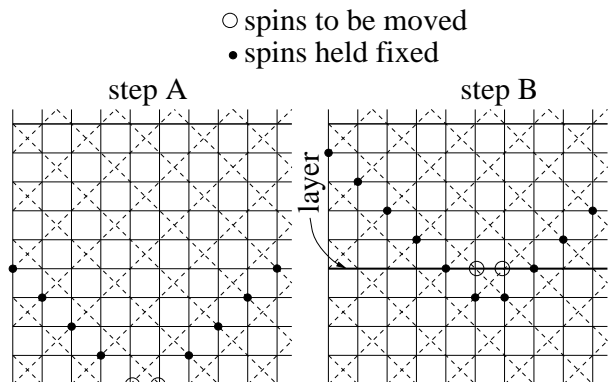,width=8cm}}
\caption{When altering the orientation of 
the spins represented by empty circles, only
spins inside the wedge defined by the fixed spins (filled circles)
have to be adjusted. For the pyrochlore lattice, a cone replaces the
wedge.}
\label{fig:cone}
\end{figure}

We make use of the following facts, which follow straightforwardly
from consideration of the ground-state configurations illustrated in
Fig.~\ref{fig:singlegs}:

(I) For a single unit of four spins, the manifold of ground states is
connected: $a$\ and $\phi$ (Fig.~\ref{fig:singlegs}), which provide a
complete parametrisation of the internal degrees of freedom of the
ground-state manifold, can be chosen independently from intervals of
the real axis. Moreover, the orientation of any two chosen spins can
be changed arbitrarily and continuously without leaving the
groundstate manifold, as long as the other two spins are
unconstrained.

(II) If two spins in a unit are antialigned, so will be the other two,
whose common axis can then be rotated arbitrarily and continuously
while keeping the orientation of the first pair fixed and remaining in
the ground-state manifold.

(III) It is possible to change the orientation of one spin in a unit
arbitrarily and continuously while that of a second is held fixed,
without leaving the ground-state manifold, as long as the other two
spins are unconstrained.  This is a special case of (I).

(IV) From II and III it follows that one can continuously change the
orientation of a pair of spins belonging to a unit in the bottom
layer, or a pair of antiparallel spins in a higher layer, and remain
within the ground-state manifold of the whole system, whilst keeping
fixed the spins that lie outside a wedge (or cone) for the square
lattice with crossings (or the pyrochlore lattice), as depicted in
Fig.~\ref{fig:cone}.

As a consequence, we can again work through the lattice layer by
layer.  The procedure is as follows.

(A) Align the spins in the bottom layer, unit by unit, to coincide with
the reference state. This fixes the two spins of each unit in the
bottom layer to be antiparallel.
As we adjust the orientation of spins in the lower
layer of each unit,
those in the upper layer of the same unit can, by I,
be reorientated to keep the unit always in a ground state.
At the same time, by IV, spins in higher layers can also be reorientated
to keep the system as a whole
in a ground state.

At this stage, spins
in the bottom layer are in their reference state.

(B) Spins in the second layer
now form antiparallel pairs,
since they belong to units which have antiparallel pairs
in the lowest layer.
The spins in the second layer can therefore, by II, be
adjusted to coincide with those in the reference state. 
While this is done, by IV, spins in higher layers
can be concurrently
reorientated to keep the system within a ground state, as in step A.

From the way we chose the reference state, it now follows that
all neighbouring spins in the second layer are pairwise
antiparallel. Therefore, we can repeat step B for the third and all
higher layers. Once we have 
done so,
all spins in the system coincide with 
those in the reference
state. Since one can go between two arbitrary ground states
via the reference state, this completes our proof
that the ground state manifold is connected.

\subsection{Pyrochlore antiferromagnets with general $n\geq2$}
\label{subsect:xygs}
The arguments presented in the previous section for Heisenberg
antiferromagnets generalise directly to $n$-component spins with
$n>3$. For the ground-state construction, the main difference is that
spins 2 and 3 in Fig.~\ref{fig:singlegs} can now be rotated in $n-2$\
directions, so that the extensive part of $D$\ is $N(n-2)$, as
expected from Maxwellian counting. 

The construction of ground states for the $XY$ model is much simpler
than for the Heisenberg model: as in the case of a Heisenberg
antiferromagnet on the kagome lattice, for a generic ground state the
number of constraints is equal to the total number of degrees of
freedom.  The construction of the ground states therefore involves
fewer choices. It proceeds as follows.  The orientations of spins in
the bottom layer can be chosen arbitrarily. The ground-state
configurations of spins in the next layer are then almost completely
determined, since each tetrahedron has two pairs of antiparallel
spins. The only freedom remaining is the discrete choice of which spin
to place on which of the two sites of the unit in the next layer,
unless the two spins of a unit in the lower layer happen to be
antiparallel.  In this case the orientation of the pair (which has to
be antiparallel) in the top layer be chosen freely.

The proof of the connectedness of the
ground-state manifold, presented above, follows essentially from the
connectedness of the ground-state manifold of a single unit and from
the fact that for any orientation of a pair of spins in a unit, the
other pair can be chosen so that the total spin of the unit vanishes.
This, along with the other steps, carries over to the case $n\geq2$.

\subsection{Ground-state correlations of the
Heisenberg pyrochlore antiferromagnet}
\label{sect:neutronexpts}

It is clear in our construction of ground states
that
spin correlations are not propagated
efficiently.  In this section, we show that, nonetheless, a few
long-range correlations in high-symmetry directions are built into the
ground states of the pyrochlore Heisenberg antiferromagnet.
We discuss the signatures of these correlations in
magnetic neutron diffraction.

\subsubsection{Correlations between planes}
\label{subsect:pyrocorrelations}

From the ground-state condition, $\bst_\alpha=0$, it follows that the
sum of all the spin vectors in two adjacent (100) planes is zero
(this sum is also the sum of $\bst_\alpha$ over tetrahedra
making up the two planes).
Therefore,
adjacent planes are antiferromagnetically correlated.
Since these correlations are long ranged, we
expect {\em sharp} peaks in the neutron scattering cross section in
the $\left<200\right>$\ directions.

These peaks differ from Bragg peaks in two ways.
First, their amplitude scales differently with sample size.
Consider a sample in the form of a cube
of side $L$, and let the total magnetisation of a (100) plane be $M$.
In a typical ground state, $M \sim L$ and the peak scattering amplitude varies
as $M\cdot L \sim L^2$, in contrast to $L^3$ for a Bragg peak.
Second, they are sharp in only one direction in reciprocal space.
Consider scattering at a point displaced from $(200)$ by the vector 
$(q_{\parallel},{\bf q}_{\perp})$. We argue that, at fixed ${\bf q}_{\perp}$,
the scattering amplitude as a function of $q_{\parallel}$
has a peak centered on $q_{\parallel}=0$, of width
$\delta q_{\parallel} \propto |{\bf q}_{\perp}|$.
Contours of constant scattering intensity therefore have a distinctive
bow-tie shape. To understand in detail the reason for this,
it is necessary to examine the correlation in the magnetisation, $\cal{M}$,
of a region of a $(100)$ plane with linear size $\cal{L}$,
and that of its equivalent, displaced by a distance $z$ in the $[100]$
direction. Let $\delta {\cal M}_z$ be the difference between these magnetisations, and measure $z$ in units of the plane spacing.
The magnetisation difference for adjacent planes, $\delta {\cal M}_1$,
arises entirely from spins belonging to tetrahedra that are only partially included in the region.
The number of such spins is proportional to the size of
the boundary of the region -- and hence to  $\cal{L}$.
Since there are only weak correlations between
individual spins on distances larger than the size of a tetrahedron,
we obtain $\delta M_1 \propto \sqrt{\cal{L}}$.
Increasing $z$, $\delta {\cal M}_z$ follows a random walk:
$\delta {\cal M}_z \propto \sqrt{z} \sqrt{\cal{L}}$. 
Since $\cal{M} \sim \cal{L}$, we obtain a correlation length
$\xi \propto \cal{L}$, and therefore 
$\delta q_{\parallel} \propto |{\bf q}_{\perp}|$.

Similarly, there are long-range correlations in the $\left<111\right>$\
directions. The $\lbrace 111\rbrace$\ planes are alternately kagome
and triangular planes. Adjacent kagome planes contain the bases of
adjacent tetrahedra, which, in the
intervening triangular planes, share a common
apex. In any ground-state,
the total magnetic moments of all (100) 
kagome planes are equal, and also opposite to 
the total magnetic moment of all (100) triangular planes.

\subsubsection{Consequences for neutron scattering experiments}
\label{subsect:elasticneutrons}

\begin{figure}
\centerline{\psfig{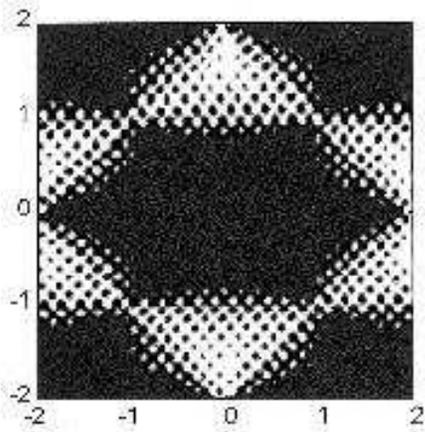}}
\caption{Static neutron scattering cross section in the $[hhl]$~plane
for a Heisenberg pyrochlore antiferromagnet, obtained numerically by
quenching the system into a ground state. Horizontal and vertical
coordinates are  $l$\  and $h$\ respectively.
Light shaded areas represent high intensity, and black
represents zero intensity.}\label{fig:martinmcneutron}
\end{figure}

In recent neutron scattering experiments
on a single crystal
sample of $CsNiCrF_6$ by
Harris~\etal,\cite{harrisneutron1,martin,martinmc}
the angular dependence
of the neutron scattering cross section is studied.  
The correlation length is longest in the $\left<100\right>$\
direction, shortest in the $\left<110\right>$\ direction, and
intermediate in the $\left<111\right>$\ direction.
Since the presence
of two species of magnetic ions in $CsNiCrF_6$ makes 
a detailed comparison with
theory difficult, Harris~\etal~also report Monte Carlo studies
\cite{martin,martinmc} of the angular neutron scattering cross section
in the $\left[ hhl \right]$\ plane,
shown in Fig.~\ref{fig:martinmcneutron}, which is taken from 
Ref \onlinecite{martin}.

The scattering is broad in most directions except the
$[100]$\ and $[111]$- directions, where narrow necks appear
at low
temperature. The scattering near these necks has the
appearance of a bow-tie, as
described above.  In addition to these bow-ties, which are narrow in
the direction parallel to the wavevector transfer, there are
subsidiary bow ties, narrow in a perpendicular direction.
Their origin can also be
explained using arguments of the kind described above.

\subsection{Local zero modes}
\label{sect:loops}

We next discuss the existence of degrees of freedom in the ground
state which involve only 
spins in a finite region in the
bulk of the system. These we call local zero modes.
They are of interest because 
unhindered rotation of finite numbers of spins
is likely to be particularly effective at destroying correlations, both
in space and in time.

In the kagome magnet, the nature of the zero modes was established in
Refs.~\onlinecite{chalkerkagome} and \onlinecite{ritchey}.  The
requirement that $\bst = 0$ leads the three spins in each triangle to
be coplanar and at relative angles of $120^\circ$.  If all the spins
on the lattice are coplanar, there are three spin orientations,
($A,B,C$), each of which occurs in each triangle once.  A zero mode,
called a weathervane defect,\cite{ritchey} arises as follows. In a
region enclosed by a line of spins of one type (say, $C$), a line of
spins (alternating between type $A$~and $B$) can be rotated about the
spin direction of $C$~at no cost in energy. Starting from a particular
state, {\em any} ground state can be constructed using these zero
modes.\cite{chalkerkagome,shender}

There is a closely related way of describing the zero modes for the $XY$
magnet on the pyrochlore lattice. They are again
associated with closed loops, this time of anti\-ferromagnetically
oriented spins. Our construction of a ground state in section
\ref{sect:gscons}~amounts to finding a set of lines through the lattice
sites with adjacent sites on a line occupied by antiparallel nearest
neighbour spins, and each site belonging to exactly one line.
Each line can be labelled with an angle giving the
orientation of its spins, and a zero mode
involves changing one such angle.
Villain\cite{villain} used a description of this kind to generate a
subset of the ground states of the Heisenberg pyrochlore
antiferromagnet. By contrast, for the $XY$ model, this approach
generates {\em all} the ground states. 

For the Heisenberg antiferromagnet on the pyrochlore lattice, the
nature of the zero modes is much more complicated, reflecting the
larger freedom in the ground-state manifold.  We have not been
able to find a simple description of a generic zero mode.
We nevertheless believe that in a sufficiently large
region of a generic ground state, there are local zero modes.
Our argument rests on counting degrees of freedom:
the number of degrees of freedom a region contributes
to the ground state is proportional to its volume, 
while fixing surrounding spins imposes
a  number
of constraints proportional to its surface.
For a large enough volume, the number of
degrees of freedom exceedes the number of such constraints.
The existence of such local modes in all ground states
is however not guaranteed. For
instance, for a state in which all spins in a given $\left( 001
\right)$\ plane are parallel, and antiparallel to the spins in
neighbouring planes, the sum rules discussed in
section~\ref{sect:neutronexpts} preclude the existence of local zero
modes.

\label{subsubsect:loops}


\section{Ground-state selection at low temperatures: analytical results}
\label{sect:singletet}

In this section, we examine the circumstances under which thermal
fluctuations induce order in geometrically frustrated
antiferromagnets.  This phenomenon -- known as order by disorder --
has been discussed in great detail for the Heisenberg antiferromagnet
on the kagome lattice, where thermal fluctuations induce coplanar
ordering of the
spins.\cite{chalkerkagome,ritchey,shender}
There is evidence from past simulations that order by disorder is not
a universal occurrence in geometrically frustrated systems -- both
Heisenberg spins on the pyrochlore lattice \cite{reimersmc,martin} and
four-component spins on the kagome lattice \cite{rute} apparently
remain disordered at low temperature -- but the systematics have not
previously been studied.

The experimental situation for pyrochlore magnets is rather
complicated. A few compounds develop long-range order at low
temperatures,\cite{fef3} whereas others undergo a spin-glass
transition.\cite{gingrascrit} It is unclear what the 
importance is of various features of real systems.
Nevertheless, we restrict our attention in the following 
to the theoretically idealised problem
of a classical antiferromagnet, without anisotropy, disorder,
further-neighbour 
or dipolar interactions.

This material is arranged as follows.  First, we consider the
analytically accessible problem of four Heisenberg or $XY$ spins on a 
single
tetrahedron. Next, we investigate the general case of
a lattice built from groups of $q$\
spins, each with $n$\ components, and ask whether thermal fluctuations
restrict the spins to an ordered (e.g.  collinear or coplanar)
configuration. Then, in Sect.~\ref{sect:gsseln}, we present the results of
numerical simulations, which test the conclusions reached from
our analytical
arguments.

\subsection{The single tetrahedron}
\label{subs:singletet}

We first study a problem simple enough to allow explicit
evaluation of some of the quantities of interest: antiferromagnetically
coupled spins occupying
the corners of an
isolated tetrahedron. 

\begin{figure}
\centerline{\psfig{figure=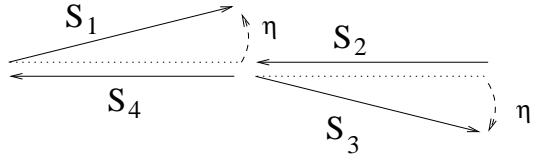,width=7cm}}
\caption{Soft fluctuations around a collinear state} 
\label{fig:softmode}
\end{figure}

There are eight degrees of freedom associated with
four Heisenberg spins, and for the system to be in a ground state, three constraints must be satisfied, since $\bst = 0$.
The ground state manifold therefore has five dimensions:
of these, three arise from global rotations, while the remaining two
can be parameterised as discussed in Sect.~\ref{subsubsect:singletetheis}.
The energy cost for fluctuations 
from this ground-state manifold in the remaining
three directions in configuration space
is, for a generic ground state, quadratic in displacement.
By contrast, for the special ground states in which spins are collinear,
energy varies quadratically with displacement from the ground state
manifold only in two directions, and quartically in the third,
indicated schematically in Fig.~\ref{fig:softmode}.
The collinear states are the obvious candidates for selection
by thermal fluctuations.

To study such selection, we have calculated the probability
distribution, $W(\theta)$, for the angle, $\theta$, between a pair of
Heisenberg spins, integrating over all orientations of the four spins
with a Boltzmann distribution and the Hamitonian of
Eq. \ref{eq:spinhamil2}.  Two factors contribute to
$W(\theta)d\theta$: the measure, $\sin(\theta)d\theta$, and a
statistical weight.  The low-temperature limit of the latter is
$[2\cos(\theta/2)]^{-1}$, the divergence as $\theta \to \pi$
reflecting the lower free energy attached to fluctuations around the
collinear state.  Combining both factors,
$W(\theta)=\sin(\theta/2)d\theta/2$: configurations which are nearly
collinear have higher weight than others in this distribution, but the
entire ground-state manifold is accessible even in the low-temperature
limit, and there is no fluctuation-induced ground-state selection.

To illustrate the alternative, consider the same problem for $XY$\ spins.
In this case, the measure contributes simply $d\theta$\ to the
distribution, $W(\theta)d\theta$, while the statistical weight
at temperature $T \ll J$ is proportional to $|\sin(\theta)|^{-1}$
for $\sin^2(\theta) \gg T/J$, and to $(J/T)^{1/2}$ 
for $\sin^2(\theta) \ll T/J$.
As a result, the weight in the limit $T\rightarrow 0$\ is 
overwhelmingly
concentrated near collinear spin arrangements ($\theta =0$ and $\pi$),
reflecting selection of these states by thermal
fluctuations. Order by disorder is just such a concentration of
statistical weight on a submanifold of ground states. Note that it can
occur in a finite system (in this case, a system of four spins), and
is quite different from the order that appears in a symmetry-breaking
phase transition, which is restricted to the thermodynamic limit.

\begin{figure}
\vspace{-2.5cm}
\centerline{\psfig{figure=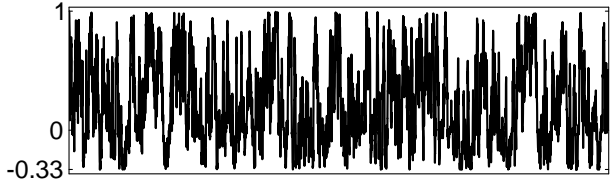,width=8cm}}
\vspace{-2.5cm}
\caption{Evolution of the collinearity parameter with Monte-Carlo time
for a single tetrahedron at $T = 2.5 \times 10^{-5}J$ for $2\times
10^7$\ Monte-Carlo steps per spin.}
\label{fig:singlecoll}
\end{figure}

It is straightforward to demonstrate these effects in Monte Carlo
simulations.  In Fig.~\ref{fig:singlecoll}, we plot the collinearity
parameter, $P(1)$ (defined in section \ref{subs:collin},
Eq.~\ref{eq:pofr}), as a function of Monte Carlo time, for a simulation
of four Heisenberg spins arranged in a single tetrahedron at the
temperature $T=2.5\times10^{-5}J$.  For the current purposes, it is
sufficient to note that the collinearity parameter takes on values
between $-1/3$~(for a state with all spins at relative angles of $
70.5 \de$\ or $109.5 \de$) and $+1$~(when all spins are collinear). We
see from Fig.~\ref{fig:singlecoll} that the system explores all ground
states, attaining values of the collinearity within $2 \times
10^{-4}$\ of the extremal ones, and is not trapped near a collinear
state.  The average of the collinearity parameter, $0.193 \pm 0.02$,
is distinct from $0$, its value in the high-temperature limit, and
close to the exact low-temperature value, $1/5$, obtained from the
expression for $W(\theta)$\ given above.

\subsection{The general problem: groups of $q$ spins with $n$
components}
\label{sect:qnspins}

We now examine whether thermal fluctuations select particular ground
states for the general class of system introduced in
Sect.~\ref{sect:models}, in which a lattice is built from $N$
corner-sharing units of $q$ spins, each having $n$ components.  We
have argued elsewhere \cite{pyrosh} that low-temperature behaviour is
characterised by a probability distribution over the ground-state
manifold, defined in the limit $T \to 0$. Let ${\bf x}$ be coordinates
on the ground-state manifold; at each point ${\bf x}$, one can
introduce local coordinates, ${\bf y}$, spanning the remaining
directions in configuration space.  Generically, the energy of the
system relative to its ground-state value will have a Taylor expansion
with the leading term
\bea
H \approx H_2 =\sum_l \epsilon_l({\bf x})y_l^2,
\label{H}
\eea 
resulting in a ground-state probability density
\bea
Z({\bf x}) \propto \int d\{y_l\} e^{-\beta H_2} 
\propto \prod_l (k_{{\rm B}}T/\epsilon_l({\bf x}))^{1/2}.
\label{Z}
\eea
In principle, order might arise in either of two ways.
First, it can happen that certain, special ground states have 
soft fluctuations,
so that at some point, ${\bf x}_0$, on the ground state manifold
(or, more generally, on
some subspace),
some of the $\epsilon_l({\bf x}_0)$ 
vanish. Then $Z({\bf x})$ will diverge as ${\bf x}$ 
approaches ${\bf x}_0$.
If any such divergences are non-integrable,
one should keep higher order terms from Eq.\ \ref{H}
when calculating $Z({\bf x})$. The result of doing so will be,
in the limit $T \to 0$, a
distribution concentrated exclusively
on the subset of ground states for which $Z({\bf x})$
is divergent: these are the configurations selected by 
thermal fluctuations.
It is this mechanism for fluctuation-induced order that
we study. There is, however, also a second possibility, 
which we do not persue here: 
it might happen that the probability density, $Z({\bf x})$,
is spread smoothly over the ground-state manifold, but that 
there nevertheless exist correlation functions which, 
when averaged with this weight, are long-ranged.

To decide whether ground states with soft modes are 
selected, it is necessary to know
the number, $M$, of $\epsilon_l$ that vanish, and the 
dimension, $S$, of
the subspace on which this happens. 
Close to this subspace, we separate
${\bf x}\equiv ({\bf u},{\bf v})$ into an 
$S$-dimensional component 
${\bf u}$, lying within the subspace, and a 
$(D-S)$-dimensional component
${\bf v}$, locally orthogonal to it, with magnitude $v$.
We expect at small $v$ the behaviour
$\epsilon_l({\bf x}) \propto v^2$ for $M$ of the $\epsilon_l$'s.
Hence, $Z({\bf x})$ diverges as $v^{-M}$ for small $v$, 
and the subspace is selected \cite{pyrosh}
as $T \to 0$ if the integral
\bea
\int Z({\bf u},{\bf v}) d{\bf v}\, \propto \,\int v^{D-S-M-1} dv
\label{eq:Z}
\eea
is divergent at small $v$. 

\begin{figure}
\centerline{\psfig{figure=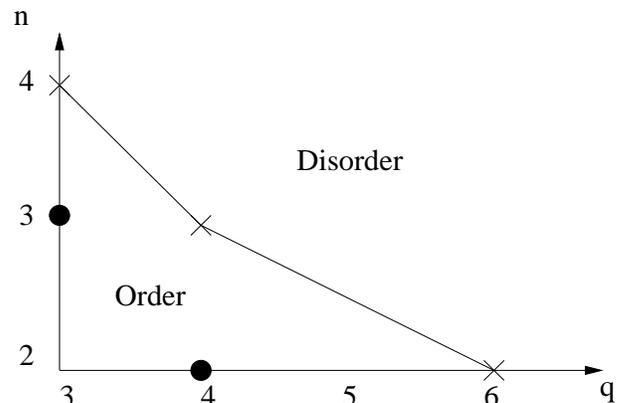,width=8cm}}
\caption{
The occurence of order by disorder for $n$-component spins arranged
in corner-sharing units, each consisting of $q$ spins. Ordered
(marginal) models are denoted by circles (crosses).
}
\label{fig:obdophase}
\end{figure}

We therefore need to consider candidate ordering patterns, and
determine the sign of $D-S-M$ in each case.  It seems in general that
the preferred ground states are ones in which spins are collinear or
coplanar, because these have the largest number of soft modes.
Collinear spin order is possible on lattices built from units
containing an even number of sites, $q$: in practice, those
constructed from tetrahedra.  Such order results in one soft mode per
unit, as illustrated in Fig \ref{fig:softmode}.  The number of soft
modes is therefore $M=N$, and (since $S=n-1$) we expect order only if
$D < N$.  Estimating $D$ as $F-K=N[q(n-1)-2n]/2$, we predict order if
$n < (q+2)/(q-2)$, and disorder if $n > (q+2)/(q-2)$.  Thus, for
pyrochlore antiferromagnets, two-component spins order, and
four-component spins do not.  The approach reaches no conclusion in
the marginal case of three-component spins, but simulations
(Refs.~\onlinecite{reimersmc,martin} and as described below) indicate
that Heisenberg spins do not order.  On lattices made from
corner-sharing triangles, such as the kagome lattice, there are no
collinear ground states; instead, coplanar order may occur.  Such
order results in $M=N/2$ soft modes;\cite{chalkerkagome} using again
the estimate $F-K=N[q(n-1)-2n]/2$ for $D$, we predict order in this
case if $n<4$ and disorder if $n>4$. Simulations of kagome
antiferromagnets demonstrate that there is indeed coplanar order for
$n=3$,\cite{chalkerkagome} and that the marginal case, $n=4$, is
disordered.\cite{rute}

Summarising, the only cases in which there is order by disorder are
$q=4,\ n=2$\ (the $XY$ pyrochlore model) and $q=3,\ n=3$\ (the
Heisenberg kagome model). In both instances, there is a low entropic
cost to enter the ordered state ($D_M=0$) and a high entropic gain
from soft fluctuations because in the ordered state the constraints,
$\bst_\alpha=0$, are not independent.
These
conclusions are depicted in Fig.~\ref{fig:obdophase}.


\section{Ground-state selection at low temperatures: numerical results}
\label{sect:gsseln}

In the following two subsections, we present the results of the Monte
Carlo simulations of $XY$ and Heisenberg pyrochlore antiferromagnets.
One aim is to test our prediction of collinear ordering for $XY$
spins. We also consider the Heisenberg model in detail, to show that
order by disorder is indeed absent in this case.  Our studies of the
Heisenberg antiferromagnet are a continuation of Reimers' pioneering
simulations \cite{reimersmc} and Zinkin's subsequent
work.\cite{martin} Our conclusions are in agreement with these
authors, in particular with the earlier -- albeit tentative -- ideas
of Zinkin, but our results are more extensive.  Reimers work
concentrated on the temperature range $T \geq 0.05J$: many of the
observations described in the following are very hard to discern or
absent in this regime.

Our simulations were carried out on systems of sizes ranging from one
unit cell ($N=2$~tetrahedra, $\ns=4$~spins) to $17^3$~unit cells
($N=9826, \ns=19652$). As pointed out in section \ref{sect:singletet},
small systems display large fluctuations, and therefore require very
long simulation runs. For $\ns = 4$, the longest simulation was $2 \times
10^8$~Monte Carlo steps per spin at $T=5 \times 10^{-5} J$. For the
largest system, however, only $1.5 \times 10^6$\ Monte Carlo steps per
spin were necessary even at the lowest temperature.

\subsection{Correlation functions}
\label{subs:collin}

In this subsection, we consider two-spin correlations
and also a correlation function which quantifies
directly the collinearity of
the spin system. In the next subsection, we discuss the heat capacity,
which is an indirect probe of the state of the system, but in
some ways more conclusive, since it is sensitive to the presence of soft
fluctuations irrespective of the type of ordering with which they are associated. 

\begin{figure}
\centerline{\psfig{figure=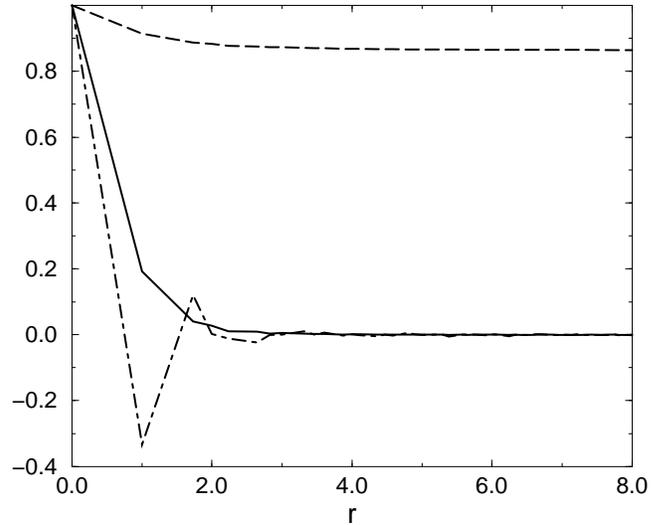,width=10cm}}
\caption{Correlation functions for the Heisenberg and
$XY$ antiferromagnets at a temperature of $T=5\times 10^{-4}J$. 
The two-spin correlation function, $Q(r)$, (dot-dashed line) 
and the collinearity correlation function, $P(r)$, (solid line) 
for a system of 2048
Heisenberg spins; and $P(r)$\ for a system of 864 $XY$ spins (dashed
line).}
\label{fig:auto}
\end{figure}

First, we demonstrate that the Heisenberg model does not have N\'eel
order, even at low temperature.  The correlation function $Q(r)
\equiv\left<\bs(0)\cdot\bs(r)\right>$\ is shown in
Fig.~\ref{fig:auto}: correlations are very small beyond the second
neighbour distance.  Second, to measure the collinearity of spins, we
evaluate the correlation function (for $n$-component spins) \bea
P(r)\equiv \frac{n}{n-1}\left(\langle\left( \bs(0) \cdot
\bs(r)\right)^2\rangle -\frac{1}{n} \right), 
\label{eq:pofr}
\eea which is constructed
to have the values $P=0$\ at infinite temperature and $P=1$\ in a
collinear state.  $P(r)$\ is shown in Fig.~\ref{fig:auto}, with $r$\
in units of nearest-neighbour distances.  The correlations for
Heisenberg spins again have a range of only two nearest neigbour
distances: there is no fluctuation-induced order.  Equally, the
predicted collinear order for $XY$ spins is confirmed: there is
long-range order in $P(r)$ at this temperature.  Note that, despite
the very low temperature, the order parameter, $P(r\rightarrow
\infty)\simeq 0.86$, is appreciably less than its maximum possible
value of 1.  We expect on general grounds that such nematic order
should be established via a first-order phase transition, but have not
attmepted to check this in detail in our simulations.

\begin{figure}
\centerline{\psfig{figure=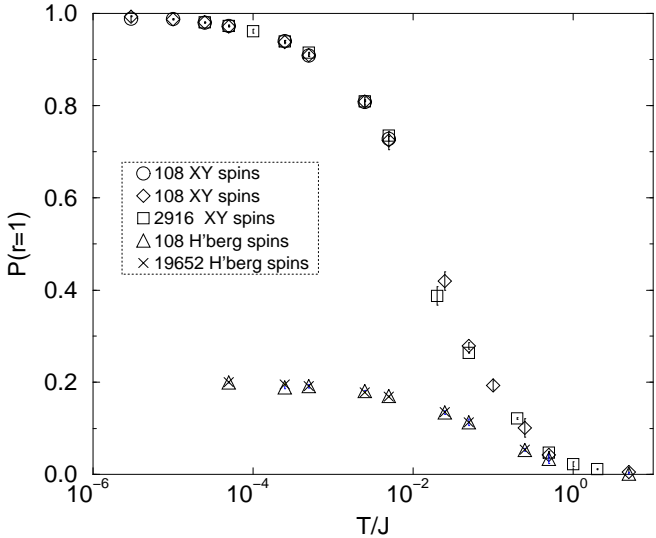,width=10cm}}
\caption{The temperature dependence of $P(r=1)$ for $XY$ and
Heisenberg spins. Most error bars are smaller than the symbols.  
Simulations started from
random initial spin configurations,
except those for the 
data points marked with open circles, which
started from collinear
spin configurations.}
\label{fig:collxy}
\end{figure}

\begin{figure}
\centerline{\psfig{figure=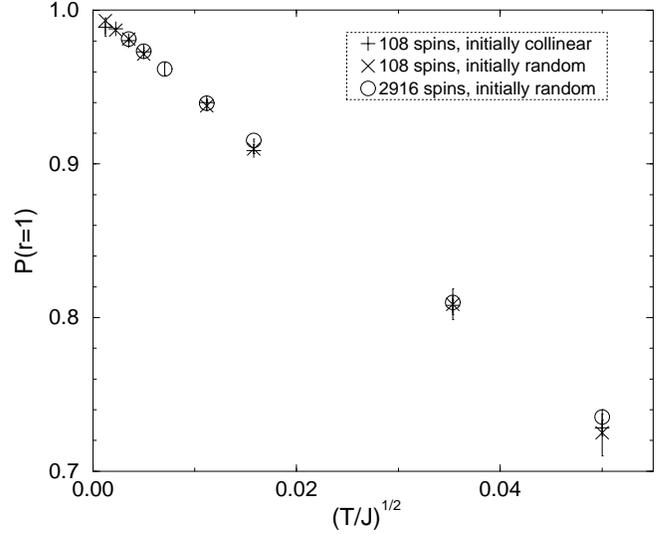,width=10cm}}
\caption{$P(r=1)$\ versus $(T/J)^{1/2}$\ for $XY$\ spins.}
\label{fig:devfromone}
\end{figure}

The temperature dependence of collinearity for neighbouring spins is
shown in Fig.~\ref{fig:collxy}. Neighbouring 
Heisenberg spins have a limiting
low-temperature value, $P(1)\simeq 0.2$, which is non-zero because the
correlation length, though small, is itself finite.  By contrast, $XY$
spins become perfectly collinear in the low-temperature limit.  The
low-temperature variation of $[1-P(1)]$, the deviation of collinearity
from its maximal value is characteristic of fluctuation-induced order.
\cite{chalkerkagome} Specifically, we expect $1-P \propto \sqrt{T/J}$
at low temperatures, because quartic modes give rise to the dominant
fluctuations at low temperatures. These modes, with coordinates
$\eta$, characterised schematically in Fig.~\ref{fig:softmode}, have
$\eta \propto T^{1/4}$ by equipartition. Since $(\bs_1 \cdot \bs_2)^2
\sim (1-\eta^2/2)^2 \sim 1-\eta^2$, we obtain $1-P(1) \propto
T^{1/2}$.  We show in Fig.~\ref{fig:devfromone} that $P(1)$ does
indeed behave in the expected way.

\begin{figure}
\centerline{\psfig{figure=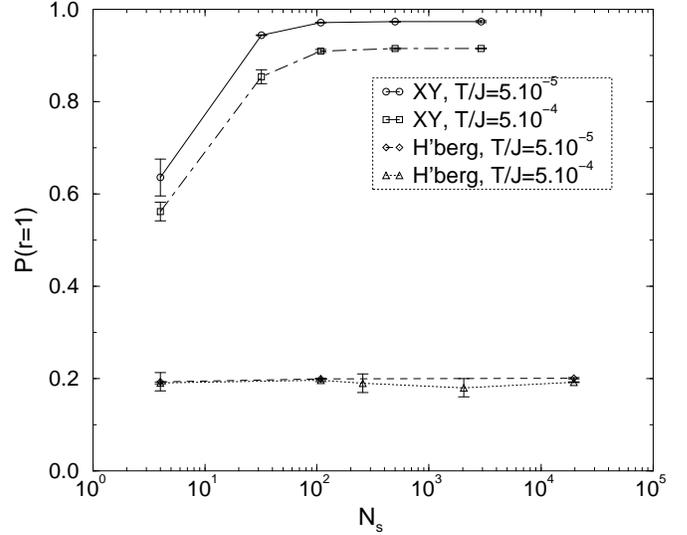,width=10cm}}
\caption{System size dependence of the collinearity parameter
$P(r=1)$\ for $XY$ and Heisenberg spins at $T=5\times 10^{-5}J$\ and
$T=5\times 10^{-4}J$.}
\label{fig:collsizedep}
\end{figure}

We have checked the dependence of our results on length of 
simulation run and system size.
To test whether the system is properly equilibrated during our 
Monte Carlo runs,
we investigate the dependence of data on initial conditions,
comparing results from random and collinear initial states.
For Heisenberg spins,  our simulations are long enough that
neither of the correlation functions studied
retains memory of the initial state.
For $XY$ spins, we are able to equilibrate $P(r)$
(see Fig.~\ref{fig:collxy}), but not $Q(r)$:
collinear order presumably hinders relaxation of two-spin correlations.
To test for finite-size effects, we carry out simulations on
systems ranging in size from $N_s=4$\ to $N_s=19652$\ spins.  Only for
small systems of $XY$-spins are marked finite size effects observed,
as shown in Fig.~\ref{fig:collsizedep}.

\subsection{Specific heat}
\label{subs:specheat}

An unbiased way to search for soft fluctuations 
and -- by implication --
fluctuation-induced order is to measure heat 
capacity.\cite{chalkerkagome}
At low temperature, one expects to be able to 
describe fluctuations of the system from a ground 
state in terms of canonical coordinates which are 
almost independent of each other. From the 
classical equipartition theorem, 
a canonical coordinate, $p$, which appears in the Hamiltonian as
$(p/p_0)^{2r}$, contributes $k_B /(2r)$\ to the
heat capacity. Hence, each quadratic mode contributes $k_B /2$, 
and each
quartic mode $k_B /4$, whereas zero modes do not contribute at
all. Determining the heat capacity therefore allows a determination of
the number of quadratic and quartic modes present.

If, as predicted, thermal fluctuations select a collinear state for
the $XY$ model, then there is one quartic and one quadratic mode per
tetrahedron. This results in a heat capacity per spin, $C$, of
$3/8 k_B$. In the absence of order, all modes are
quadratic, and $C=k_B/2$. For the Heisenberg antiferromagnet, there
are four degrees of freedom per tetrahedron, one of which is a zero
mode. If there is no order, we expect $C=3/4 k_B$; if a
collinear state is selected by thermal fluctuations, $C=5/8 k_B$.
In finite-sized systems, the heat capacity per spin is reduced.
For our choice of periodic boundary conditions, this manifests itself
in the correction 
{$C(N)=\left( {(N-1)/{N}} \right)
C(\infty ) $.}

\begin{figure}
\centerline{\psfig{figure=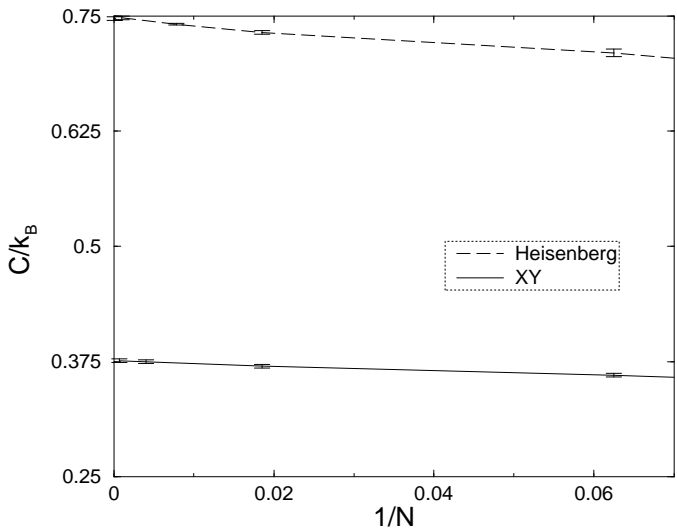,width=10cm}}
\caption{The specific heat for different system sizes for the $XY$ and
the Heisenberg antiferromagnet on the pyrochlore lattice. The lines
are a guide to the eye.
}
\label{fig:specheatsystsize}
\end{figure}

As shown in Fig.~\ref{fig:specheatsystsize}, we find in the limit $N
\rightarrow \infty$, that $C=0.376 \pm 0.002$\ for $XY$ spins, and
that $C=0.747 \pm 0.002$\ for Heisenberg spins.  This is consistent
with the presence of order for $XY$ spins, with one quartic mode per
tetrahedron, as expected. For Heisenberg spins, we obtain an upper
limit of 0.04 quartic modes per tetrahedron.

\section{The dynamics of the pyrochlore antiferromagnet}
\label{chapter:dy}

We now turn to time-dependent correlation functions, and ask how the 
system explores the vicinity of
its ground-state manifold at low temperature. We
study the spin autocorrelation function with precessional dynamics,
both analytically and by numerical integration of the equations of 
motion. 
Two of the facts established in the previous sections
have important implications for spin dynamics.
First, we have shown that the ground-state manifold is
connected, which means that the magnet does not get trapped in
a particular state at low temperatures by internal energy
barriers. Second, we have shown that there is no entropic selection
of special ground states at low temperatures, which
suggests that the dynamics is unlikely to be hindered by free energy
barriers. We indeed find that spin correlations relax relatively 
rapidly
even at low temperatures, having a time-scale that diverges
only as $T^{-1}$, and not, for example, according to an
Arrhenius law. 

\subsection{Derivation of an effective spin dynamics}
\label{sect:effequation}

Consider, in the first instance, the 
dynamics linearised around a ground state. In this approximation, 
a system of
$N$ tetrahedra (hence with $2N$ spins and a $4N$-dimensional phase-space) 
will have $2N$ normal modes.
Some of these modes will be conventional, finite frequency spin-waves, 
but a fraction will have zero frequency, because there is no restoring 
force for displacements in phase space that take the system from one 
ground state to another. Beyond the harmonic approximation, non-linear 
terms in the full equations of motion will have various consequences: 
the conventional, finite-frequency modes will acquire a finite 
lifetime; and coupling between these modes and the ground-state 
coordinates will drive the system around its ground-state manifold.
We find that there are three distinct time-scales at low temperature.
The period of the highest frequency spin-waves, $O(\hbar/J)$, 
sets the shortest scale; their lifetime, $\tau_s \sim \hbar/[Jk_BT]^{1/2}$, 
provides an intermediate scale;
while the longest scale is the decay time of the autocorrelation 
function, $\tau \sim \hbar/k_BT$. This separation of time-scales greatly 
simplifies the problem.

Our starting point is the equation of motion,
\bea
\frac{d\bs_i}{dt}=
\bs_i\times {\bf H}_i(t)
\equiv -J\,\bs_i\times({\bf L}_{\alpha}+{\bf L}_{\beta}),
\label{eqnofmotion}
\eea
where we have set $\hbar=1$. ${\bf H}_i(t)$ is the exchange field 
acting at site $i$, which
can be expressed in terms of
${\bf L}_{\alpha}$ and ${\bf L}_{\beta}$, the total spins
of the two tetrahedra to which $\bs_i$ belongs.
Summing over the sites of a tetrahedron, the time-dependence of 
${\bf L}_{\alpha}$ is
\bea
\frac{d{\bf L}_{\alpha}}{dt}=
-J\,\sum_{\beta}\bs_{\alpha\beta}\times{\bf L}_{\beta},
\label{eqnofmotion2}
\eea
in which the notation $\bs_{\alpha\beta}$ has been introduced for 
the spin common to the tetrahedra $\alpha$ 
and $\beta$. 

\begin{figure}
\centerline{\psfig{figure=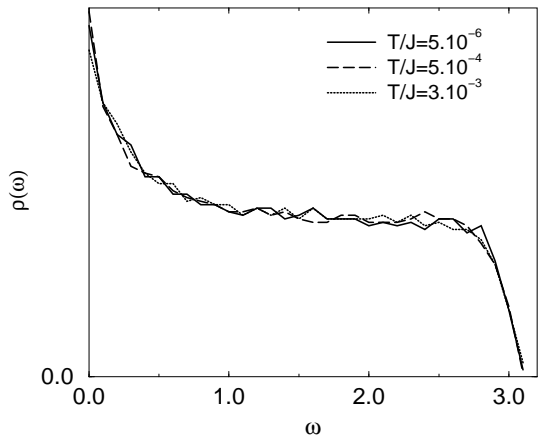,width=8cm}}
\caption{The density of states, $\rho(\omega)$ for a system of 2048
Heisenberg spins at low temperature.}
\label{rho}
\end{figure}

The right side of Eq.\ \ref{eqnofmotion2} implicitly defines a $3N
\times 3N$ matrix, ${\bf M}$, acting on a vector constructed from the
components of the ${\bf L}_{\alpha}$'s. This matrix, being real and
antisymmetric, has eigenvalues which are purely imaginary and occur in
pairs, $\pm i \omega$, related by complex conjugation. For a
ground-state spin configuration, the magnitudes of these eigenvalues
are the frequencies of $3N/2$ normal modes, while the real and
imaginary parts of the associated eigenvectors are canonically
conjugate coordinates for the modes.  The remaining $N$ directions in
phase space are spanned by coordinates having ${\bf L}_{\alpha}=0$ for
all $\alpha$, and therefore lie within the ground-state manifold.  The
matrix ${\bf M}$ is well-defined and has purely imaginary eigenvalues
for any spin configuration; for a low-temperature spin configuration,
the eigenvalue magnitudes are presumably good approximations to the
normal mode frequencies in a nearby ground state. We display in
Fig.~\ref{rho} the density of states, $\rho(\omega)$, on a linear
scale, obtained by diagonalising ${\bf M}$ for low-temperature
pyrochlore spin configurations generated in a Monte Carlo
simulation. It is noteworthy that $\rho(\omega)$ appears to be finite
at $\omega = 0$: $\rho(\omega)$ neither includes a divergent
contribution, proportional to $\delta(\omega)$, nor does it vanish as
$\omega \to 0$.

The fact that $\rho(\omega)$ does not contain a delta 
function
at $\omega=0$ gives information on how canonically conjugate 
pairs of coordinates appear in the linearised dynamics. Quite 
generally, the Hamiltonian in the harmonic approximation can be 
reduced to the form
\bea
H= \sum_{l=1}^{2N}(\alpha_lp_l^2 + \beta_lq_l^2),
\eea
where $p_l$ and $q_l$ are a  canonically conjugate pair 
of
coordinates.
We know from the ground-state construction described earlier 
that $1/4$ of these coordinates belong to the ground-state 
manifold, and therefore that $1/4$ of the $4N$ numbers 
$\{ \alpha_l, \beta_l\}$ are zero. Oscillations of the 
coordinates $p_l,q_l$ have zero frequency if 
$\alpha_l \cdot \beta_l = 0$,
and so the fraction of zero-frequency spin-wave modes might, 
in principle, range from $1/4$, if for each of these modes both 
$\alpha_l=0$ and $\beta_l=0$, to $1/2$, if for each of these modes 
only one of $\alpha_l$ and $\beta_l$ is zero.
The fact that $\rho(\omega)$ does not contain a delta function at 
$\omega=0$
implies that all (except for a fraction vanishing in the thermodynamic 
limit)
of the $3N/2$ modes derived from ${\bf M}$ have non-zero frequency, and 
therefore that only the remaining $1/4$ of modes have zero frequency. 
Hence, for every $l$, either {\it both} $\alpha_l$ and $\beta_l$ are
zero or {\it neither} is zero: coordinates in the ground-state 
manifold all appear in canonically conjugate pairs. 

The non-zero density of states apparent at small frequency in
Fig.~\ref{rho} is in striking contrast to the behaviour, $\rho(\omega)
\propto \omega^2$, that occurs in both N\'eel-ordered antiferromagnets
(including ordered states of the pyrochlore antiferromagnet
\cite{lacroix}) and conventional spin-glasses.  The arguments used by
Halperin and Saslow \cite{halperin-saslow} and by Ginzburg
\cite{ginzburg} to predict propagating long-wavelength modes in
spin-glasses depend on the ground state having a stiffness.  This
stiffness appears to be missing in ground states of the pyrochlore
antiferromagnet, because of the many zero modes.

The exchange field, ${\bf H}_i(t)$, appearing in the equation of
motion, Eq.\ \ref{eqnofmotion}, can be written as a superposition of
contributions arising from the finite-frequency modes, in terms of
vectors, ${\bf u}_l(i)$, determined by $\bf M$, and amplitudes, $A_l$,
determined by the initial conditions: \bea {\bf H}_i(t)=\sum_l A_l
{\bf u}_l(i) e^{i \omega_l t} + c.c.\,.  \eea In the harmonic
approximation, the amplitudes, $A_l$, are time-independent, but in the
full dynamics their magnitude and phase will change on a time-scale
which defines the spin-wave lifetime, $\tau_s$. We postpone detailed
discussion of the temperature dependence of $\tau_s$\ until the end of
this section, but note that, at low temperature, $\tau_s$ is large
compared to the typical spinwave period, $J^{-1}$. 

If the equation of
motion is integrated over time-intervals longer than $\tau_s$,
contributions from modes with frequencies $\omega_l \gg \tau_s^{-1}$
average to zero, while those from modes with $\omega_l \alt
\tau_s^{-1}$ fluctuate randomly, according to the time-dependence of
$A_l(t)$.  Hence ${\bf H}_i(t)$ has mean value zero, and has
fluctuations which are characterised most importantly by their
low-frequency spectral density, \bea \int_{-\infty}^{\infty}dt' \langle
{\bf H}_i(t) \cdot {\bf H}_i(t') \rangle\, \equiv\, 2\Gamma\,.
\label{D}
\eea
In terms of the amplitudes, $A_l(t)$,
\bea
\Gamma = \sum_l \int \langle A_l(0) A_l^*(t) \rangle e^{i \omega_l t}
|{\bf u}_l(i)|^2 dt\,.
\label{G2}
\eea
Since, for large $\tau_s$, only the low-frequency modes 
contribute to $\Gamma$,
and since, from equipartition,  $\langle|{\bf H}_i(t)|^2\rangle \sim J^2 
\langle |{\bf L}_{\alpha}|^2 \rangle \sim Jk_{\rm B}T$, 
we have $\Gamma \propto Jk_{\rm B}T\rho(0^+)$. Note that, 
as $\rho(\omega) \propto J^{-1}$, $\Gamma$ is independent of $J$.

We now proceed to
calculate long-time spin correlations from Eq.\ \ref{eqnofmotion}, 
by
treating ${\bf H}_i(t)$ as Gaussian white noise with the 
correlator
\bea
\langle {\bf H}_i(t) \cdot {\bf H}_i(t')  \rangle = 2\Gamma \delta(t-t')\,.
\label{corr}
\eea
To do so involves several assumptions. The most important of 
these is that the spinwave lifetime, $\tau_s$, which sets the 
width in time of the delta function in Eq\ \ref{corr},
is small compared to the decay time of the spin 
autocorrelation function. 
We show below that this is asymptotically exact as $T/J \to 0$. 
Further, in taking $\Gamma$ to be a constant, rather than a functional 
of the instantaneous spin configuration, we implicitly neglect 
variations with spin configuration in the local density of states,
$ \sum_l |{\bf u}_l(i)|^2 \delta(\omega_l)$. Such fluctuations 
certainly exist, but seem from our numerical studies only to 
have a small effect on the form of the spin autocorrelation function.
Solving the Langevin equation that results from 
treating ${\bf H}_i(t)$ in Eq.\ \ref{eqnofmotion} as white noise, 
we obtain $\langle \bs_i(0) \cdot \bs_i(t) \rangle = e^{-\Gamma t}$ 
and hence (reinstating $\hbar$)
\bea
\langle \bs_i(0) \cdot \bs_i(t) \rangle = \exp(-ck_{\rm B}Tt/\hbar)\,,
\label{auto}
\eea
where $c$ is a dimensionless constant of order unity. Therefore, the
autocorrelation time $\tau=\Gamma^{-1}=\hbar/(ck_BT)$. We emphasise 
again that,
at $T \ll J/k_{\rm B}$, it is $T$ alone, and not $J$, which sets 
the scale for long-time dynamics.

To complete this discussion, it is necessary to estimate the 
spinwave lifetime, $\tau_s$. There are two physical processes 
that contribute to $\tau_s$. One, common to all antiferromagnets, 
is the anharmonic interaction between different finite-frequency 
modes, which here results in a lifetime varying as $T^{-1}$ for 
small $T$. It is, however, overwhelmed by a second process, specific 
to systems with many ground-state degrees of freedom, in which 
finite-frequency modes are mixed by the motion of the system 
between different ground states. More formally, on time 
scales $\gg \hbar/J$, the matrix ${\bf M}$ is time-dependent. 
The linearised equations of motion, with time-dependent ${\bf M}$, 
define an autonomous dynamical problem in which the instantaneous 
normal mode amplitudes, $A_l(t)$, are time-dependent. The 
time-dependence of the matrix elements of ${\bf M}$ mixes 
amplitude, initially concentrated in a single mode, $l$, over 
all modes lying within a window of frequencies around $\omega_l$. 
From time-dependent perturbation theory, a fractional change, $f$, 
in matrix elements spreads amplitude over a frequency window whose 
width, $\delta \omega$, forms a fraction $f$ of the entire spinwave 
spectrum, so that $\delta \omega \sim fJ/\hbar$. And from our 
results for the spin autocorrelation function, the fractional 
change in matrix elements during a time-interval $\tau_s$ 
is $f=k_{\rm B} T \tau_s/\hbar$. The spinwave lifetime is the 
time at which the frequency window resulting from this $f$ 
has width $\delta \omega \sim 1/\tau_s$, and so we obtain 
$\tau_s \sim \hbar/[J k_{\rm B} T]^{1/2}$. As required for the 
consistency of our arguments, at low temperatures this is 
indeed a much shorter timescale than that for the decay of 
the spin autocorrelation function.

As the temperature is raised towards $T\sim J$, this separation of
timescales breaks down. The precession on the previously shortest
timescale then becomes visible, and the autocorrelation decays
initially as $1-A(t)\propto t^2$\ rather than $1-A(t)\propto
t$. This is indeed observed  in our numerical simulations described in
the next section (Fig.~\ref{fig:zerop}).

\subsection{Molecular dynamics simulations}
\label{sect:molecdyn}

In this subsection, we present results obtained from numerical
simulations of the dynamics of the Heisenberg pyrochlore
antiferromagnet. In these simulations we evaluate
the autocorrelation function
\beq
A(t) \equiv \left< \bs_i(0) \cdot \bs_i(t)\right>.
\label{eq:autocor}
\eeq 
We find that $A(t)$\ decays exponentially
in time. The simulations confirm the predictions of the preceding section, namely that the timescale for the
dynamics, $\tau$, varies as $T^{-1}$.
Finally, we do not discover any sign of spin freezing
even at temperatures as low as $T=5\times 10^{-4}J$.

We generate uncorrelated, thermalised initial configurations by Monte
Carlo simulation, from which the equation of motion, Eq.\
\ref{eqnofmotion}, is integrated using a fourth-order Runge-Kutta
algorithm.  Related calculations for the kagome Heisenberg
antiferromagnet have been described previously by Keren.\cite{keren}

Some details of our procedure are as follows. We choose the
integration time-step so that energy is conserved to at least one part
in $10^{8}$.  The temperature range of the simulations covers three
orders of magnitude, the lowest temperature being $T=5\times
10^{-4}J$.  The system sizes studied range from 32 spins to 2048
spins. There are marked finite size effects in the smaller systems,
which we believe result from all spins precessing together about the
total magnetisation, ${\bf M}_{tot}$, of the system.  Since $|{\bf
M}_{tot}|^2 \sim N$, the precession rate varies as $N^{-1/2}$, and
decreases rather slowly with increasing system size.  For the results
presented, we hasten convergence to the thermodynamic limit by adding
the term $J{\bf M}_{tot}^2$\ to the Hamiltonian, which constrains the
total magnetisation to be independent of system size and near zero.
As a result, values of the decay time, $\tau$, coincide for systems
with 500 and 2048 spins.

\subsubsection{The functional form of $A(t)$}
\label{subsect:formofa}

\begin{figure}
\vspace{-1cm}
\centerline{\psfig{figure=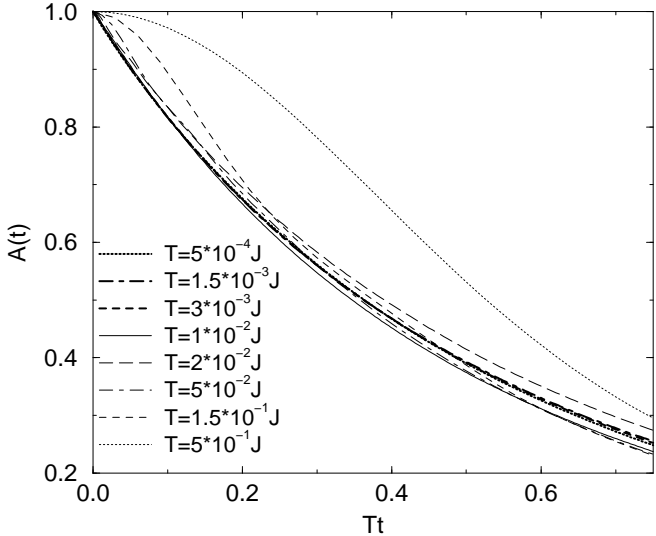,width=10cm}}
\caption{The autocorrelation function as a function of the rescaled
time, $Tt$.}
\label{fig:zerop}
\end{figure}

\begin{figure}
\vspace{-1cm}
\centerline{\psfig{figure=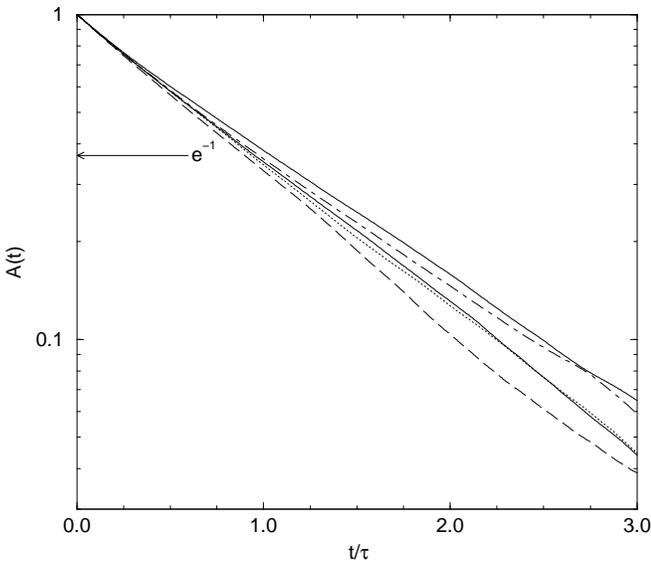,width=10cm}}
\caption{The decay of $A(t)$\  on a logarithmic scale for five
different runs at the same 
temperature, $T/J=6\times 10^{-3}$, in a system of 2048 spins.}
\label{fig:error}
\end{figure}

From the analytic calculation presented in subsection \ref{sect:effequation},
we expect $A(t)$\ to depend on time and temperature only through
the combined variable, $Tt$.
We show in Fig.~\ref{fig:zerop} $A(t)$ as a function of this 
scaling variable,
at various temperatures and over
one and a half decay times, $\tau$, for a system of 2048
spins.
The collapse of the data onto a single curve, at all except the highest temperatures ($T/J \geq 0.1$), is striking evidence in 
support of our analytic results.
To demonstrate the accuracy with which the decay of $A(t)$\ 
at low temperature is exponential, and to indicate the
magnitude of finite-size effects in our results,
we show in Fig~\ref{fig:error} data for $A(t)$ on a logarithmic scale,
for runs starting from different initial configurations 
generated at the same temperature.

\begin{figure}
\centerline{\psfig{figure=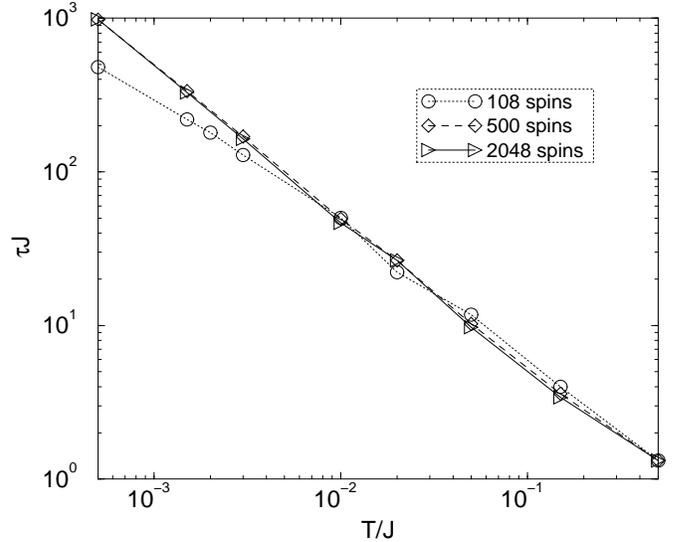,width=10cm}}
\caption{The decay time as a function of tempeature for three
different system sizes.
}
\label{fig:tjp0}
\end{figure}

To examine quantitatively the temperature
dependence of the decay time, $\tau$, we fit data at each temperature
to an exponential, $\exp\left( -t/\tau\right)$.
The resulting values for $\tau$\ are displayed in
Fig.~\ref{fig:tjp0}.
In order to extract the temperature dependence of $\tau$, we fit it to the
power law $\tau = {\cal {A}} T^{-\zeta}$.\cite{pyrosh} 
Excluding temperatures $T/J\geq 0.15$,
we obtain $\zeta = 0.998\pm 0.012$\ and ${\cal{A}}=0.53\pm
0.04$.  This result agrees with and confirms our prediction that
$\zeta=1$.

\subsection{Inelastic neutron scattering}
\label{subsect:dynexpts}

Inelastic neutron scattering provides the most detailed probe of
dynamical correlations. We expect from the results described above
that diffuse inelastic scattering in the temperature range $T_F <T \ll
|\Theta_{CW}|$ should have a Lorentzian lineshape in energy, with a
width, $\Gamma$, varying as $\Gamma=ck_BT$, where $c$ is of order
unity. Although we have not explicitly examined the dependence of
dynamic correlations on wavevector, $k$, it seems likely that main
contribution to the inelastic linewidth in the low-temperature limit
should be roughly wavevector independent. One reason for thinking this
is that the decay of the autocorrelation function probably arises
because of rotation of relatively small clusters of spins - the local
ground-state degrees of freedom identified in \ref{sect:loops}.  In
consequence, we expect dynamical correlations to be short-ranged in
space and broad in wavevector.  In addition, conservation laws which
might result in significantly different behaviour, for example $\Gamma
\propto k^2$ for small $k$ from conservation of spin density, do not
appear to be in operation: since the magnetisations of individual
tetrahedra are identically zero in classical ground states,
instantaneous magnetisation fluctuations can decay without spreading
to large distances. Thermally induced fluctuations in the
magnetisations of tetrahedra may result in an additional, diffusive
component to spin correlations, with an amplitude that vanishes in the
low-temperature limit.

Inelastic neutron scattering from the pyrochlore antiferromagnet
$CsNiCrF_6$, which has $|\Theta_{CW}|\simeq 70$K and $T_F\simeq
2.2$K,\cite{martin} is reported in Ref.~\onlinecite{harrisneutron2},
in which the energy dependence of scattering is fitted by a Lorentzian. The linewidth decreases
as temperature is decreased below 70 K, but the data seem
insufficiently precise to test whether $\Gamma$ is linear in
$T$. Similar experiments on $SCGO$, in which $|\Theta_{CW}|$\ is
around 500~K and $T_F\simeq 3.5$K, yield a Lorentzian inelastic
lineshape with a temperature dependence of the width 
which is encouragingly close
to linear, and of order $k_BT$, over the temperature range from $30K$
to $290K$.\cite{broholm1,aeppli}
 
\section{Concluding remarks}

We have presented a detailed theoretical analysis of the low-temperature
properties of a class of geometrically frustrated classical
antiferromagnets, with particular emphasis on the pyrochlore
Heisenberg antiferromagnet, which has a macroscopically
degenerate ground-state manifold.  Given the connectedness of this
ground-state manifold and the absence of appreciable free energy
barriers, there seems to be no mechanism for localising the system in
a particular region of the ground-state manifold, and it is therefore
unlikely that the spin glass transition observed in most experiments
is a feature of the disorder-free classical isotropic
Heisenberg pyrochlore antiferromagnet. Rather, we find correlation
functions to be short-ranged in both space and time, and conclude that
the spins continue to fluctuate strongly down to the lowest
temperatures.

We have analysed for the first time the low-energy dynamics of these
geometrically frustrated antiferromagnets.  Our discussion does not
depend on details of the pyrochlore lattice structure.  In fact, we
expect it also to apply to the Heisenberg model defined on the
$SCGO$-lattice of Fig.~\ref{fig:scgo}, since that model has $D_M/N>0$\
and does not -- following the arguments presented in
Ref.~\onlinecite{pyrosh} and Sect.~\ref{sect:singletet} -- display
order by disorder.  These properties are in striking contrast to those
of the kagome Heisenberg antiferromagnet, and $SCGO$\ is much more
similar to a pyrochlore magnet than to the kagome system.

We expect our results to be robust against the introduction of a
small concentration of vacancies,\cite{villain} certainly provided the
average defect spacing is larger than extent of the most local ground-state
degree of freedom in the pure system. 
Behaviour characteristic of the pure system
may persist to much
higher defect concentrations, since $D_M/N>0$\ as long as more than
three quarters of all sites are occupied.

\section*{acknowledgements}
We are particularly grateful for discussions with M. Zinkin, whose
experiments and simulations, described in Ref.~\onlinecite{martin},
stimulated this work.  We also thank G. Aeppli, P. Chandra, R. Cowley,
M. Harris and P. Holdsworth for helpful discussions. We acknowledge
the hospitality of the Institute for Theoretical Physics, UCSB, during
the course of part of the work.  It was supported in part through
EPSRC Grant GR/J8327 and NSF Grant PHY94-07194.

\end{document}